\documentclass[aps,preprint]{revtex4}
\usepackage{graphicx}
\begin{document}
\title{\Large \bf Self-Dual Fields Harbored by a Kerr-Taub-bolt Instanton}
\author{\large A. N. Aliev}
\affiliation{ Feza G\"ursey Institute, P.K. 6  \c Cengelk\" oy,
81220 Istanbul, Turkey}
\author{\large Cihan Sa\c{c}l\i o\u{g}lu}

\affiliation{Faculty of Engineering and Natural Sciences, Sabanci
University, 81474 Tuzla, Istanbul, Turkey}
\date{\today}
\begin{abstract}

We present a new exact solution for self-dual Abelian  gauge fields
living on the space of the Kerr-Taub-bolt instanton, which is a
generalized example of asymptotically flat instantons with
non-self-dual curvature, by constructing the corresponding square
integrable harmonic form on this space.

\end{abstract}
\maketitle

\newpage


Gravitational instantons are usually defined as complete nonsingular
solutions of the vacuum Einstein field equations in Euclidean space
\cite{h}, \cite{gh}, \cite{yau}-\cite{hit}. Among other things, they
play an important role in the path-integral formulation of quantum
gravity \cite{h1}, \cite{GibPerry} forming a privileged class of
stationary phase metrics that provide the dominant contribution to
the path integral and mediate tunneling phenomena between
topologically  inequivalent vacua. The first examples of
gravitational instanton metrics were obtained by complexifying  the
Schwarzschild, Kerr and Taub-NUT spacetimes through analytically
continuing them to the Euclidean sector \cite{h},\cite{gh}. The
Euclidean Schwarzschild and Euclidean Kerr solutions do not have
self-dual curvature though they are asymptotically flat at spatial
infinity and periodic in imaginary time, while the Taub-NUT
instanton is self-dual. However, there exists another type of
Taub-NUT instanton which, unlike the first one, is not self-dual and
possesses an event horizon ("bolt") \cite{page}. The generalization
of this Taub-bolt metric to the rotating case was given in
\cite{gperry}.

Another class of gravitational instanton solutions consists of the
Eguchi-Hanson metric \cite{eh} and the multi-centre metrics of
\cite{gh}, which include the former as a special case. These metrics
are asymptotically locally Euclidean with self-dual curvature and
admit a hyper-K\"ahler structure. (For a review see \cite{egh}). The
hyper-K\"ahler structure of gravitational instantons and some
properties of gravitational instantons which are derivable from
minimal surfaces in $3$-dimensional Euclidean space were examined in
\cite{an1}, \cite{an2} using the Newman-Penrose formalism for
Euclidean signature.

A fundamental difference between manifolds that have  Euclidean $(
++++)$ and  Lorentzian $(-+++) $ signatures is that the former can
harbor self-dual gauge fields that have no effect on the metric,
while in the latter external fields serve as source terms in field
equations. In other words, since the energy-momentum tensor vanishes
identically for self-dual gauge fields, solutions of Einstein's
equations automatically satisfy the system of coupled
Einstein-Maxwell and Einstein-Yang-Mills equations. The
corresponding self-dual gauge fields are inherent in the given
instanton metric. Furthermore, in Euclidean signature, Weyl spinors
also have vanishing energy-momentum tensor and vector and axial-
vector bilinear covariants. Hence they cannot appear as source terms
in the field equations as well. The explicit solutions for different
configurations of some "stowaway" gauge fields and spinors living on
well-known Euclidean-signature manifolds have been obtained in a
number of papers (see \cite{HawkPope}-\cite{tekin}) .

In recent years, motivated by Sen's $S$-duality conjecture
\cite{sen}, there has been some renewed interest in self-dual gauge
fields living on well-known Euclidean-signature manifolds. The gauge
fields were studied by constructing self-dual square integrable
harmonic forms on given spaces. For instance, the square integrable
harmonic $2$-form in self-dual Taub-NUT metrics was constructed in
\cite{gibbons}, its generalization to the case of complete
noncompact hyper-K\"ahler spaces was given in \cite{hitchin}.
However the similar square integrable harmonic form on manifolds
with non-self-dual metrics was found only for the simple case of the
Euclidean-Schwarzschild instanton \cite{etesi}. In this note we
shall give a new exact solution to describe the Abelian "stowaway"
gauge fields harbored by the Kerr-Taub-bolt instanton, which is a
generalized example of asymptotically flat instantons with
non-self-dual curvature. This is achieved  by explicit construction
of the corresponding square integrable harmonic form on the space.

The Euclidean Kerr-Taub-bolt instanton was discovered by Gibbons and
Perry \cite{gperry} as a rotating generalization of the earlier
Taub-bolt solution \cite{page} with non-self-dual curvature. This
Ricci-flat metric is still asymptotically flat and in
Boyer-Lindquist type coordinates it has the form

\begin{equation}
ds^{2}=\Xi \left( \frac{dr^{2}}{\Delta }+d\theta ^{2}\right)
+\frac{\sin
^{2}\theta }{\Xi }\left( \alpha \,dt+P_{r}\,d\varphi \right) ^{2}+\frac{\Delta }{%
\Xi }\left( dt+P_{\theta }\,d\varphi \right) ^{2}\,\,, \label{ktb}
\end{equation}
where the metric functions are given by
\begin{eqnarray}
\Delta &=& r^{2}-2Mr-\alpha ^{2}+N^{2}\,,\\
\Xi &= & P_{r}-\alpha P_{\theta }=r^{2}-(N+\alpha \cos \theta
)^{2}\,\,, \\
P_{r}&=&r^{2}-\alpha ^{2}-\frac{N^{4}}{N^{2}-\alpha ^{2}}\,\,\,,\\
P_{\theta }& = & -\alpha \sin ^{2}\theta +2N\cos \theta
-\frac{\alpha N^{2}}{N^{2}-\alpha ^{2}}\,\,.
\end{eqnarray}
The parameters $\, M ,\,\, N ,\,\, \alpha \,\,$ represent the
"electric" mass, "magnetic" mass and "rotation" of the instanton,
respectively.

When $\,\alpha=0\,$ this metric reduces to the Taub-bolt instanton
solution found in \cite{page} with an event horizon and
non-self-dual curvature. If $\,N=0\,$, we have the Euclidean Kerr
metric. Thus one can say that the metric (\ref{ktb})  generalizes
the Taub-bolt solution of \cite{page} in same manner just as the
Kerr metric  generalizes the Schwarzschild solution. The coordinate
$\,t\,$ in the metric behaves like an angular variable and  in order
to have a complete nonsingular manifold at values of $\,r\,$ defined
by equation $\,\Delta=0\,$ , $\,t\,$ must have a period
$\,2\pi/\kappa\,$. The coordinate $\,\varphi\,$ must also be
periodic with period $\,2\pi \,(1-\Omega /\kappa )\, $, where the
"surface gravity" $\,\kappa\,$ and the "angular velocity" of
rotation $\,\Omega,$ are defined as
\begin{eqnarray}
\kappa & = & \frac{r_{+}-r_{-}}{2\,r_{0}^{2}}\,\,,~~~~~~~ \Omega
=\frac{\alpha }{r_{0}^{2}}\,\,, \label{sgav}
\end{eqnarray}
with
\begin{eqnarray}
r_{\pm }&=& M \pm \sqrt{M^{2}-N^{2}+\alpha ^{2}}\,\,\,,~~~~~~~
r_{0}^{2} = r_{+}^{2}-\alpha ^{2}-\frac{N^{4}}{N^{2}-\alpha
^{2}}\,\,\,.
\end{eqnarray}
As a result one finds that the condition
$$\kappa =\frac{1}{4\mid N\mid }$$ along with $ \Xi \geq 0 \,$ for $\,r>r_{+}\,$ and $\,0\leq \theta \leq \pi \,$
guarantees that $\,r=r_{+}\,$  is a regular bolt in the nonsingular
manifold of (\ref{ktb}) .

We shall also need the basis one-forms for the metric (\ref{ktb})
which can be chosen as
\begin{eqnarray}
e^{1} &=&\left(\frac{\Xi }{\Delta
}\right)^{1/2}dr\,\,,~~~~~e^{2}=\Xi ^{1/2}d\theta \,,\nonumber
\\[2mm]
e^{3} &=&\frac{\sin \theta }{\Xi ^{1/2}}\left(\alpha \,dt+P_{r}\,d\varphi\right)\,\,,
\\[2mm]
e^{4} &=&\left(\frac{\Delta }{\Xi }\right)^{1/2}(dt+P_{\theta
}\,d\varphi )\,\,.\nonumber \label{bforms}
\end{eqnarray}

The isometry properties of the Kerr-Taub-bolt instanton with respect
to a $\, U(1)\,$- action in imaginary time imply the existence of
the Killing vector field
\begin{eqnarray}
\frac{\partial}{\partial
t}&=&\xi^{\mu}_{(t)}\,\frac{\partial}{\partial x^{\mu}}\,\,.
\label{killing}
\end{eqnarray}
We recall that the fixed point sets of this Killing vector field
describe a two-surface, or bolt, in the metric. We shall use the
Killing vector to construct a square integrable harmonic $2$-form on
the Kerr-Taub-bolt space. It is well-known that for a Ricci-flat
metric a Killing vector can serve as a vector potential for
associated Maxwell fields in this metric \cite{papa}. Since  our
Kerr-Taub-bolt instanton is also Ricci-flat, it is a good strategy
to start with the Killing one-form field
\begin{equation}
 \xi = {\xi}_{(t) \mu}\,d\,x^\mu\,
\end{equation}
which is obtained by lowering the index of the Killing vector field
in (\ref{killing}). Taking the exterior derivative of the one-form
in the metric (\ref{ktb}) we have
\begin{eqnarray}
\label{2form}
 d\xi &=&\frac{2}{\Xi^2}\left\{ \,\left[M r^2 +
\left(\alpha M \cos\theta-2 N r+M N\right)\left( N+\alpha \cos\theta
\right)\right] e^{1} \,\wedge e^{4} \right. \\[2mm]  & & \left. \nonumber
 -\left[N\,\left(\Delta + \alpha^2 +\alpha^2\, \cos^2 \theta\right) + 2\,
\alpha (N^2-M r) \cos\theta\right] e^{2} \,\wedge e^{3}\,\right\}\,.
\end{eqnarray}
In this expression we have used the basis one-forms (\ref{bforms})
in order to facilitate the calculation of its Hodge dual, which is
based on the simple relations
\begin{eqnarray}
^{\star}\left(e^{1} \,\wedge e^{4}\right)&=& e^{2} \,\wedge
e^{3}\,\,,~~~~ ^{\star}\left(e^{2} \,\wedge e^{3}\right)= e^{1}
\,\wedge e^{4}\,\,.~~~~~~~~~ \label{duals}
\end{eqnarray}
Straightforward calculations using the above expressions show that
the two-form (\ref{2form}) is both closed and co-closed, that is, it
is a harmonic form. However the Kerr-Taub-bolt instanton does not
admit hyper-K\"ahler structure, and the two-form given by
(\ref{2form}) is not self-dual. Instead, we define the
(anti)self-dual two form
\begin{equation}
F=\frac{\lambda}{2}\,(d\xi \pm \,^{\star} d\xi)\,, \label{sdual}
\end{equation}
where $\,\lambda\,$ is an arbitrary constant related to the dyon
charges carried by the fields and  the minus sign refers to the
anti-self-dual case. Taking equations (\ref{2form}) and
(\ref{duals}) into account in this expression, we obtain  the
harmonic self-dual two-form
\begin{equation}
F=\frac{\lambda (M-N)}{\Xi^2}\,\left(r+N +\alpha \cos\theta\right)^2
\left( e^{1} \,\wedge e^{4} + e^{2} \,\wedge e^{3}\right)\,\,,
\label{sd2form}
\end{equation}
which implies the existence of the potential one-form
\begin{equation}
A=- \lambda\,(M-N)\,\left[\cos\theta\, d\varphi + \frac{r+N +\alpha
\cos\theta}{\Xi} \,\left(d t+ P_{\theta}\, d \varphi)\right)\right]
\,\,. \label{spotform}
\end{equation}
After an appropriate re-scaling of the parameter $\,\lambda\,$,
which includes the electric coupling constant as well ,
 a string singularity at $\theta =0$ or $\theta =\pi $ in this
expression is avoided as usual by demanding the familiar Dirac
magnetic-charge quantization rule.

From equation (\ref{sdual}) we also find the corresponding
anti-self-dual two-form
\begin{equation}
F=\frac{\lambda (M+N)}{\Xi^2}\,\left(r-N -\alpha \cos\theta\right)^2
\left( e^{1} \,\wedge e^{4} - e^{2} \,\wedge e^{3}\right)\,\,,
\label{asd2form}
\end{equation}
The associated potential one-form is given by
\begin{equation}
A=- \lambda\,(M+N)\,\left[-\cos\theta\, d\varphi + \frac{r-N -\alpha
\cos\theta}{\Xi} \,\left(d t+ P_{\theta}\, d \varphi)\right)\right]
\,\,. \label{aspotform}
\end{equation}

For $\,\alpha=0\, $, the above expressions describe self-dual, or
anti-self-dual Abelian gauge fields living on the space of a
Taub-Nut instanton with an horizon \cite{page}. In the absence of
the "magnetic" mass $\,(N=0)\,$ we have the gauge fields harbored by
the Euclidean-Kerr metric. The latter can also be obtained from the
potential one-form in the Kerr-Newman dyon metric after an
appropriately Euclideanizing it and setting the electric and
magnetic charges equal to each other ( see \cite{carter}) .

Next, we shall show that these self-dual and anti-self-dual harmonic
two-forms are square integrable on the Kerr-Taub-bolt space. This
can be shown by explicitly  integrating the Maxwell action. For the
self-dual two-form we have
\begin{equation}
\frac{1}{4\pi ^{2}}\int F\wedge F=\frac{\lambda^2}{2\,\pi ^{2}}\,
(M-N)\,\int_{0}^{t_0} dt \int_{0}^{\varphi _{0}}d\varphi =\frac{2\lambda ^{2}%
}{\kappa }\,(M-N)\,\left( 1-\frac{2 \,\alpha }{r_{+}-r_{-}
}\right)\,\,, \label{maxact}
\end{equation}
where  $\, t_0=2\pi/\kappa\,$ and $\,\varphi_0=2\,\pi(1-\Omega
/\kappa)\,$. Since this integral, which represents the second Chern
class $\,C_2\,$ of the $\,U(1)$-bundle, is finite, the self-dual
two-form $\,F\,$ is square integrable on the Kerr-Taub-bolt space.
For an anti-self-dual $\,F\,$, a plus sign must be introduced
between $M$ and $N$ in (\ref{maxact}).

It is also useful to calculate the total magnetic flux $\Phi $ which
is obtained by integrating the self-dual 2-form $\,F\,$ over a
closed $2$-sphere $\Sigma $  of infinite radius; dividing this by
$2\pi $ gives the first Chern class with minus sign
\begin{equation}
-C_{1}=\frac{\Phi }{2\pi }=\frac{1}{2\pi }\int_{\Sigma }F = 2\lambda
\,(M-N)\,\left( 1- \frac{2 \,\alpha }{r_{+}-r_{-} }\right)\,,
\end{equation}
which must be equal to an integer $\,n\,$ because of  the Dirac
quantization condition. We see that the periodicity of angular
coordinate in the Kerr-Taub-bolt metric affects the magnetic-charge
quantization rule in a non-linear way. It involves both  the
"electric" and "magnetic"  masses and the  "rotation" parameter.

\vspace{5mm} We would like to  thank  M. J. Perry for helpful
discussions.


\begin{thebibliography}{99}

\bibitem{h}  S. W. Hawking,   Phys. Lett. {\bf 60A} (1977) 81
\bibitem{gh}  G. W. Gibbons and S. W. Hawking,  Phys. Lett. B {\bf 78} (1978) 430
\bibitem{yau}  S. T. Yau, Comm. Pure and Appl. Math. {\bf 31} (1978) 339
\bibitem{ahs}  M. F. Atiyah, N. Hitchin  and I. M. Singer, Proc. Roy. Soc. A
{\bf 362} (1978) 425
\bibitem{hit} N. Hitchin, Polygons and Gravitons, Math. Proc. Camb. Phil. Soc.
{\bf 85} (1979) 465
\bibitem{h1} G. W. Gibbons and S. W. Hawking,  Phys. Rev. \textbf{D15 }(1977)
2752
\bibitem{GibPerry} G.W. Gibbons, S.W. Hawking and M. J. Perry, Nucl. Phys.
\textbf{B138} (1978)141
\bibitem{page} D. N. Page, Phys. Lett. 78B (1978) 249
\bibitem{gperry} G.W. Gibbons and M. J. Perry, Phys. Rev.\textbf{\ D22 }
(1980) 313
\bibitem{eh}  Eguchi T and  Hanson A J 1978 Phys. Lett. {\bf 74B} 249
\bibitem{egh} T. Eguchi, P.B. Gilkey and A.J. Hanson, Physics Reports \textbf{%
\ 66 }(1980) 213
\bibitem{an1}  A. N. Aliev  and  Y. Nutku, Class. Quant. Grav. {\bf 16} (1999) 189
\bibitem{an2}  A. N. Aliev, M. Hortacsu, J. Kalayci and  Y. Nutku, Class. Quant. Grav. {\bf 16} (1999)
631
\bibitem{HawkPope} S. W. Hawking and C. N. Pope, Phys. Lett. \textbf{73B}
(1978) 42
\bibitem{duffm} M. J. Duff and J. Madore, Phys. Rev. {\bf D18} (1978) 2788
\bibitem{Comtet} H. Boutaleb-Joutei, A. Chakrabarti and A. Comtet, Phys. Rev.
\textbf{D20} (1979) 1884, Phys. Rev. \textbf{D20} (1979) 1898, Phys.
Rev. \textbf{D21} (1980) 979, Phys. Rev. \textbf{D21} (1980) 2280,
Phys. Rev. \textbf{D21} (1979) 2285
\bibitem{Charap} J. M. Charap and M. J. Duff, Phys. Lett. \textbf{69B}
(1977) 445, Phys. Lett. \textbf{71B} (1977) 219
\bibitem{cih} C. Sa\c{c}l\i o\u{g}lu, Class. Quantum Grav. \textbf{17 }(2000)
485
\bibitem{tekin} B. Tekin, Phys. Rev. \textbf{D 65} (2002) 084035
\bibitem{sen}  Sen A, Phys. Lett. B {\bf 329} (1994) 217
\bibitem{gibbons}  G. W.  Gibbons, Phys. Lett. B {\bf 382} (1996) 53
\bibitem{hitchin} N. Hitchin, Commun. Math. Phys. {\bf 211} (2000) 153
\bibitem{etesi}  G. Etesi, J. Geom. Phys. \textbf{37} (2001) 126
\bibitem{papa} A. Papapetrou, Ann. Inst. H. Poincare {\bf 4} (1966) 83
\bibitem{carter} B. Carter, {\it in Black Holes }, eds. C. De Witt and B. De Witt,
(1973)







\end{thebibliography}
\end{document}